\documentclass[twocolumn,showpacs,preprintnumbers,amsmath,amssymb]{revtex4}

\usepackage{graphicx}
\usepackage{dcolumn}
\usepackage{bm}

\begin{document}

\preprint{to appear in Phys. Rev. B}

\title{
Unconventional magnetic transition and transport behavior in Na$_{0.75}$CoO$_2$ 
}

\author{T.~Motohashi$^1$}
 \email{t-mot@msl.titech.ac.jp} 

\author{%
R.~Ueda$^1$, E.~Naujalis$^1$, T.~Tojo$^1$, I.~Terasaki$^2$, 
T.~Atake$^1$, M.~Karppinen$^1$ 
}

\author{H.~Yamauchi$^1$}

\affiliation{%
$^1$Materials and Structures Laboratory, 
Tokyo Institute of Technology, Yokohama 226-8503, Japan \\
$^2$Department of Applied Physics, Waseda University, 
Tokyo 169-8555, Japan 
}




\date{%
Received 16 December 2002
}

\begin{abstract}
Here we report an unconventional magnetic and transport phenomenon 
in a layered cobalt oxide, Na$_x$CoO$_2$. 
Only for $x$ = 0.75, a magnetic transition of the second order 
was clearly detected at $T_m$ $\sim$ 22 K where an apparent 
specific-heat jump, an onset of extremely small spontaneous 
magnetization, and a kink in resistivity came in. 
Moreover large positive magnetoresistance effect was observed 
below $T_m$. 
These features of the transition strongly indicate the appearance 
of an unusual electronic state that may be attributed to 
the strongly-correlated electrons in Na$_{0.75}$CoO$_2$. 
\end{abstract}

\maketitle

Since the discovery of the first high-$T_c$ superconducting copper oxide, 
tremendous amount of research has been conducted on $3d$ transition metal 
oxides to search for novel functions of the strongly-correlated electron system. 
In such correlated-electron systems, spin-charge-orbital interactions are 
in a subtle balance such that various different electronic states could be 
stabilized depending on the thermodynamic conditions. 
As a consequence, unexpectedly large response could be induced against 
a slight change in, e.g., the carrier density and/or lattice distortions: 
as an example of the former serves the hole doping into insulating 
CuO$_2$ planes of layered copper oxides, which induces high-$T_c$ 
superconductivity \cite{Ginsberg89a}, and as an example of 
the latter the isovalent rare-earth ({\it RE}) substitution 
at the ({\it RE},Sr) site in ({\it RE},Sr)MnO$_3$, 
which controls the colossal magnetoresistance phenomenon \cite{Tokura99a}. 
It is thus probable that highly precise control on the chemical 
composition in a variety of correlated-electron-system materials 
yields unconventional electronic states accompanied by unexpected 
electronic/magnetic phase transitions. 

Layered cobalt oxides of the Na$_x$CoO$_2$ system have been known as 
thermoelectric materials since Terasaki {\it et al.} \cite{Terasaki97a} 
reported that some of them showed simultaneously large 
thermoelectric power ($S$) and low electrical resistivity ($\rho$). 
The large value reported for thermoelectric power of Na$_{0.5}$CoO$_2$, 
i.e. $S$ $>$ 50 $\mu$V / K at 300 K, is hardly understood within 
the framework of the conventional one-electron picture. 
It has been suggested \cite{Terasaki00a} that both the strong electron 
correlation and the cobalt spin state play a crucially important role in 
facilitating such a high value of thermoelectric power. 
The crystal structure of Na$_x$CoO$_2$ consists of a single atomic 
layer of Na ions sandwiched by two CoO$_2$ layers. 
The CoO$_2$-layer block is constructed with a two-dimensional 
triangular lattice of cobalt ions. 
For Na$_x$CoO$_2$, paramagnetic behavior was reported 
\cite{Tanaka94a,Ray99a,Ono01a} in the whole temperature range 
studied (4.2 - 300 K) without any distinct magnetic orderings. 
Recently, we established a sample-preparation method named 
``rapid heat-up'' technique that has enabled us to precisely control 
the Na content in Na$_x$CoO$_2$ samples \cite{Motohashi01a}. 
For thus prepared samples of Na$_{0.75}$CoO$_2$, 
we detected a magnetic ordering below 22 K. 

In this article, we report the magnetic transition 
observed only for samples with the solubility-limit Na content, 
i.e. $x$ = 0.75, in the layered-cobalt-oxide system, Na$_x$CoO$_2$. 
A magnetic transition of the second order was clearly confirmed at 
$T_m$ = 22 K, being accompanied by an apparent specific-heat jump, 
a small spontaneous magnetization, a kink in the $\rho - T$ curve, 
and a drastic enhancement in the positive magnetoresistance. 
In Na$_x$CoO$_2$, any long-range ordering of spins is likely to be 
suppressed by the cobalt triangular lattice that tends to possess 
geometrical frustration amongst the competing interactions. 
The magnetism of unusual characteristics observed in the present work 
in our Na$_{0.75}$CoO$_2$ samples could be an example of the recently 
high-lighted phenomena coming out of the frustration systems 
\cite{Diep94a} upon being triggered by a tiny stimulation only, 
i.e. in the present case by a precise control of the Na nonstoichiometry. 

The Na$_{0.75}$CoO$_2$ samples were synthesized employing the 
``rapid heat-up'' technique \cite{Motohashi01a}. 
By x-ray powder diffraction, the pellet samples were confirmed to be 
single phase of the hexagonal $\gamma$-Na$_x$CoO$_2$ 
\cite{Fouassier73a,Jansen74a} without any traces of impurity phases. 
Careful high-resolution TEM and ED investigations revealed that 
the samples were homogeneous and contained no anomalous 
nano-scale clusterings \cite{Motohashi02a}. 
Magnetization was measured by a SQUID magnetometer (Quantum Design; MPMS-XL). 
Magnetotransport measurements were performed using a four-point-probe 
apparatus (Quantum Design; PPMS) in a transverse-field condition, 
i.e. under a magnetic field normal to the applied current. 
For every kind of characterization, multiple number of specimens 
were measured to confirm reproducibility. 
Heat capacity was measured using a home-made adiabatic calorimeter. 
Measurements were carried out between 8 and 330 K for a powder sample of 
$\sim$ 6 g prepared at 900$^{\circ}$C by the rapid heat-up technique. 
More detailed descriptions of the present method for heat capacity 
measurements are given elsewhere \cite{Tanaka94b}. 

Figure~\ref{22K-anomalies} shows the dependences of specific heat ($C_p$), 
magnetic susceptibility ($M / H$), and electrical resistivity ($\rho$) 
of Na$_{0.75}$CoO$_2$ on temperature ($T$). 
A sharp jump at 21.8 K in the $C_p - T$ curve 
(Fig.~\ref{22K-anomalies}(a)) represents the magnetic transition 
temperature ($T_m$). 
The jump exhibits a shape typical for the second-order phase transition. 
Below $T_m$, magnetic susceptibility rapidly increases under 
an applied magnetic field of 1 to 100 Oe, exhibiting a distinct hysteresis 
between the zero-field-cooled (ZFC) and field-cooled (FC) curves 
(Fig.~\ref{22K-anomalies}(b)). 
Moreover, the resistivity gradually decreases in the lower temperature region, 
yielding a kink around $T_m$ in the $\rho - T$ curve 
(Fig.~\ref{22K-anomalies}(c)). 
Note that the $\rho - T$ curve has a steeper slope below $T_m$ 
and decreases toward a very low residual resistivity ($\rho_r$) 
of $\sim$ 15 $\mu\Omega$ cm. 
As clearly shown in Fig.~\ref{22K-anomalies}, 
all of these anomalies occur at the same temperature, i.e. at $T_m$. 

In order to estimate the baseline about the specific-heat jump 
in Fig.~\ref{22K-anomalies}(a), 
the $C_p - T$ curve was fitted using the following formula: 
$$ C_p(T) = f_D(T / \Theta_D) + \gamma T, $$
where $\Theta_D$, $\gamma$, and $f_D$ denote respectively, the Debye 
temperature, the electronic specific-heat coefficient, and Debye's formula 
for which the degree of freedom was assumed to be 3 per atom. 
Based on least-square calculations, the values of $\Theta_D$ and $\gamma$ 
were determined to be 553 K and 25.9 mJ / K$^2$ mol, respectively. 
The present value of $\Theta_D$ is larger than those reported for 
(Na,Ca)Co$_2$O$_4$ samples by Ando {\it et al.}, 
i.e. 350-390 K \cite{Ando99a}, 
but the $\gamma$ value is in good agreement with those for the same samples, 
i.e. 52-56 mJ / K$^2$ mol [(Na,Ca)Co$_2$O$_4$]. 
After subtracting the baseline (shown by the solid line 
in Fig.~\ref{22K-anomalies}(a)), 
the enthalpy and entropy changes were calculated at $\Delta H$ = 0.975 J / mol 
and $\Delta S$ = 0.0481 J / K mol, respectively. 

Figure \ref{CW-plot} shows the inverse susceptibility ($\chi^{-1}$)
vs temperature plot at $H$ = 10 Oe. 
The plot follows the Curie-Weiss law above $T_m$: 
$\chi = C / (T - \Theta)$, where $C$ and $\Theta$ denote 
the Curie constant and the Weiss temperature, respectively. 
The values for $C$ and $\Theta$ were determined to be 
0.234 emu K/mole Oe and $-166.4$ K, respectively. 
The negative value of $\Theta$ suggests an antiferromagnetic interaction 
(some negative values were also reported for $\Theta$'s of Na$_x$CoO$_2$ 
in previous works \cite{Ray99a,Ono01a}). 
This value for $C$ yields an effective moment ($\mu_{eff}$) 
of (1) 1.37 $\mu_B$ / Co site if all the cobalt atoms are assumed 
equivalent, and (2) 2.74 $\mu_B$ / Co$^{4+}$ if only Co$^{4+}$ spins 
contribute (and all other Co$^{3+}$ are in the low spin state, 
i.e. $S$ = 0). 
At least the latter case is unlikely since the theoretical 
``spin-only'' value of Co$^{4+}$ in the low spin state 
is 1.73 $\mu_B$ which is much less than 2.74 $\mu_B$. 
Nonetheless there are no other evidences to possitively support 
$\mu_{eff}$ = 1.37 $\mu_B$ for the former model. 

Figure \ref{M-H} shows the magnetization ($M$) as a function 
of the applied field ($H$) at various temperatures; 
2, 5, 10, 20, and 50 K. 
In the $M - H$ loop at 2 K, non-linear increase in magnetization 
and a narrow hysteresis are observed in the low field region. 
A non-linear $M-H$ relation is also seen at 5 and 10 K, 
though the distinct hysteresis is hardly seen. 
On the other hand, at 20 K (just below $T_m$) and 50 K (above $T_m$), 
the magnetization is linearly proportional to the applied field, 
indicating the absence of long-range order at these temperatures. 
Spontaneous magnetization, $M_s$, was estimated at $H$ = 0 
through linearly extrapolating the $M - H$ curve and plotted 
in Fig.~\ref{Ms-T} as a function of temperature. 
The magnitude of $M_s$ increases rapidly as temperature is lowered. 
However, it is as small as 10$^{-4}$ $\mu_B$ / Co site 
even at 2 K: this value is extraordinarily small in comparison 
with those of conventional ferromagnetic compounds. 

One of the most characteristic features of the present compound 
is large positive magnetoresistance (MR) effect seen in the 
magnetically ordered state. 
Figure~\ref{MR} shows the dependence on temperature of 
the degree of MR effect, as defined by 
$\Delta \rho_H / \rho_0 \equiv (\rho_H - \rho_0) / \rho_0$, 
where $\rho_H$ is the resistivity in an applied field of $H$. 
The magnitude of MR effect was below the detection limit 
($\sim$ 0.1\%) at temperatures higher than 25 K. 
With decreasing temperature, on the other hand, the $\Delta \rho_H / \rho_0$ 
value at 7 T abruptly increases and reaches up to $\sim$ 0.3 at 2 K. 
The onset temperature of the MR effect is in good agreement 
with $T_m$ = 22 K, which clearly indicates 
that the positive MR effect is triggered by the magnetic transition. 
As shown in the inset of Fig.~\ref{MR}, the degree of MR effect 
also depends on the applied magnetic field such that: 
$\Delta \rho_H / \rho_0 \propto H^2$. 

It should be pointed out that the occurrence of magnetic ordering 
was theoretically predicted for Na$_x$CoO$_2$ at $x$ = 0.5 
by Singh \cite{Singh00a}. 
He predicted that this compound possesses a large density of state 
at $E_F$ and therefore an inherent electronic instability yielding 
an itinerant ferromagnetism. 
However, the magnetic behavior of the present Na$_{0.75}$CoO$_2$ sample 
is difficult to be ascribed to conventional itinerant ferromagnetism. 
Reasons for this are as follows. 
(1) It was found that Arrott plots [$M^2$ vs $H/M$ relation] 
\cite{Wohlfarth77a} do not to form a straight line nor passes 
the origin at $T_m$. 
(2) The Weiss temperature ($\Theta$) is negative, being in contrast 
to positive $\Theta$ values for conventional ferromagnets. 
(3) The $M_s$ value does not rapidly increase starting at $T_m$ 
as temperature is lowered. 
Taking into account these facts as well as the distinct thermomagnetic 
irreversibility below $T_m$ (see Fig.~\ref{22K-anomalies}(b)), 
it is considered that in the present compound a ferromagnetic interaction 
starts to develop at $T_m$, and eventually forms a long-range order 
among the magnetic moments at lower temperatures \cite{comment2}. 

Tsukada {\it et al.} \cite{Tsukada01a} found a ferromagnetic 
ordering below 3.2 K with small $M_s$ and large negative MR 
in single-crystal samples of a Pb-doped Bi-Sr-Co-O misfit-layer compound. 
However, the magnetism in Na$_{0.75}$CoO$_2$ should be different 
from that of the (Bi,Pb)-Sr-Co-O compound, because signs of MR 
are opposite between the two compounds. 
Therefore, the key to understand the electronic state of Na$_{0.75}$CoO$_2$ 
must be lying in the nature of the positive MR effect. 
Recently, a similar positive MR effect was reported in a weak-ferromagnetic 
Co-doped FeSi alloy, and it was attributed to some quantum 
interference effects \cite{Manyala00a}. 
For Na$_{0.75}$CoO$_2$, however, the MR characteristics were much 
different from those of the (Fe,Co)Si alloy in terms of the dependences 
on temperature and field. 
We also drew a scaling plot as that given in Ref. \cite{Manyala00a}
and confirmed that our plot for Na$_{0.75}$CoO$_2$ 
does not form a single universal relation, implying totally different 
mechanisms for the two cases. 
In conventional metals the MR effect arises from the ``bending'' of 
electron trajectory by the Lorentz force such that Kohler's plot, 
i.e. $\Delta \rho_H / \rho_0$ vs $(H / \rho_0)^2$, forms a single 
universal relation that is independent of temperature \cite{Ziman63a}. 
In Fig.~\ref{Kohler}, shown are the MR data collected for the 
present Na$_{0.75}$CoO$_2$ sample at various temperatures below $T_m$. 
Obviously, all the data points in the temperature range of 2 - 15 K 
follow more or less a universal relation, i.e. the present MR data 
obey Kohler's rule. 
Therefore, we conclude that the positive MR effect of Na$_{0.75}$CoO$_2$ 
is most likely originated from the conventional orbital motion of carriers. 
Since it is known that in such a situation the degree of MR effect 
is proportional to the square of carrier mobility, the large magnitude of 
MR effect is considered to be caused by a drastic enhancement 
in carrier mobility in the magnetically ordered state. 

It is noteworthy that the observed entropy change at $T_m$, 
i.e. $\Delta S$ = 0.0481 J / K mol, is exceedingly small, 
corresponding to only $\sim$ 1\% of the theoretical value, i.e. 
$R\ln 2$ = 5.76 J / K mol for the state in which $F$ (degree of freedom) = 1 
and $S$ (spin quantum number) = 1/2, and spins are all perfectly ordered. 
On the other hand, as seen in Fig.\ref{22K-anomalies}(c), 
the resistivity slope, i.e. $d\rho / dT$, is enhanced 
by a factor of 2 below 22 K, 
implying that the Drude weight, $n / m^*$ (where $n$ is the carrier 
concentration and $m^*$ the carrier effective mass), 
has been significantly reduced. 
Such a reduction in Drude weight is considered to be 
caused by a (pseudo) gap opening. 
As a possible ordered state accompanied by the tiny entropy change 
and an open pseudogap, either a charge-density-wave (CDW) or 
a spin-density-wave (SDW) state is inferred. 
Upon assuming a state with an open pseudogap at temperatures below $T_m$, 
both the resistivity drop towards an extremely low residual resistivity 
and the large positive MR effect can be consistently explained. 
The strong dependence of resistivity on temperature 
(Fig.~\ref{22K-anomalies}(c)) 
implies that the electrical conduction is mainly 
influenced by the electron-electron scattering. 
In such a situation, the total carrier-mobility is expected to be enhanced 
if one of the Fermi surfaces that yields the lower-mobility carriers 
disappears, as resistivity is inversely proportional to 
the {\it total} scattering time averaged over the whole Fermi surfaces. 
It is considered that below $T_m$ the Fermi surface 
for the lower-mobility carriers vanishes upon a CDW or SDW formation, 
whereas the other for the higher-mobility carriers still survives 
\cite{comment1}. 
Note that this view is consistent with the calculated band picture 
by Singh \cite{Singh00a} who predicted that in Na$_{0.5}$CoO$_2$ 
there exist two Fermi surfaces of different natures: 
one is a cylindrical/hexagonal hole-surface 
originated from the localized $a_{1g}$ band, and the other 
a hole-like section with the itinerant $a_{1g} + e_g$ character. 
It is plausible that the localized $a_{1g}$ and itinerant $a_{1g} + e_g$ 
bands correspond respectively to the low-mobility and to the 
high-mobility carriers in our present model. 

The simultaneous onsets of thermomagnetic irreversibility, finite 
entropy jump, resistivity drop, and large positive MR effect imply 
that the magnetism of Na$_{0.75}$CoO$_2$ is unusual in comparison 
with any cases of classical magnetic states. 
In fact, the simultaneous enhancement in magnetization is difficult 
to understand by assuming a CDW/SDW state below $T_m$: 
this is the most mysterious feature of the present ``unconventional'' 
magnetic transition in Na$_{0.75}$CoO$_2$. 
To solve this difficulty, two distinct ordered states 
may be required to coexist below $T_m$: 
one is a CDW or SDW state, and the other is a magnetically ordered 
state which alone is mainly responsible for the magnetization anomaly. 
In such a case, cobalt species in the two different states are 
required to coexist. 
A previous NMR study \cite{Ray99a} suggested two inequivalent 
cobalt sites in a Na$_x$CoO$_2$ sample with $x$ = 0.5. 
On the other hand, judging from the magnitude of spontaneous 
magnetization of Na$_{0.75}$CoO$_2$, the magnetically ordered state 
should arise from a very small number of cobalt species, which may be 
difficult to be confirmed by any existing site-specific 
spectroscopy techniques. 
A similar behavior was reported for a Cu-substituted 
Na$_x$CoO$_2$ in which an SDW-like transition was induced by 
the Cu-for-Co substitution \cite{Terasaki02a}. 
Similarity of the two cases suggests that spatial inhomogeneity 
in the cobalt state does not drastically modify the pristine electronic 
structure: the existence of inequivalent cobalt species itself is not 
an intrinsic feature of the present magnetic transition. 

Finally, we comment on our successful observation of 
this unconventional magnetism in Na$_{0.75}$CoO$_2$. 
The key technique to have an Na$_x$CoO$_2$ sample exhibit the magnetic 
transition is to precisely control the Na nonstoichiometry. 
It is known that Na-rich Na$_x$CoO$_2$ samples are difficult to be synthesized 
with a conventional solid-state reaction method, as a large deficiency 
results in the Na content inevitably \cite{Kawata99a}. 
Using various Na$_x$CoO$_2$ samples with $x$ = 0.65 - 0.75 prepared with 
both our RH technique and a conventional method, 
{\it only the $x$ = 0.75 RH sample} showed the low-temperature 
second order transition. 
Very recently, Takeuchi {\it et al.} \cite{Takeuchi02a} studied the magnetic 
properties of their Na$_x$CoO$_2$ samples with a nominal composition 
of $x$ = 0.75 and found an irreversible magnetic behavior below 13 K, 
suggesting a short-range ferromagnetic coupling. 
However, the samples showed no indication of spontaneous magnetization 
down to 2 K, though an SG-like anomaly was seen at 3 K. 
This result implies that the magnetic property of Na$_x$CoO$_2$ 
is highly sensitive to the Na nonstoichiometry. 
The reasons why no magnetic transitions were detected for various 
Na$_x$CoO$_2$ samples of previous works were: 
(1) samples with high enough Na contents had not been synthesized, 
that is, in previous works even the starting compositions 
were rather Na-poor, e.g. $x$ = 0.50 and $x$ = 0.55 for 
Refs. \cite{Tanaka94a,Ray99a} and \cite{Ando99a}, respectively, 
and (2) conventional powder methods that most likely resulted 
in Na losses were employed. 

In summary, we have confirmed the existence of an unconventional 
magnetic and transport phenomenon in Na$_x$CoO$_2$ with a precisely 
controlled Na content, $x$ = 0.75. 
The magnetic transition with $T_m$ = 22 K was of the second order, 
being accompanied with a specific-heat jump. 
At the same temperature, $T_m$, weak spontaneous magnetization 
($M_s$ = 1.2$\times$10$^{-4}$ $\mu_B$ / Co site at 2 K), 
rapid resistivity drop towards an extremely low residual resistivity 
($\rho_r$ $\sim$ 15 $\mu\Omega$ cm), and large positive magnetoresistance 
effect ($\Delta \rho_{{\rm 7T}} / \rho_0$ $\sim$ 0.3 at 2 K) 
were found to occur simultaneously. 
All these strongly indicate the appearance of an unconventional 
electronic state stemmed from the strong correlation of electrons. 

We thanks Dr. Taniyama for his fruitful discussion on the magnetism. 
The present work was supported by a Grant-in-aid for Scientific Research 
(Contract No. 11305002) from the Ministry of Education, Culture, Sports, 
Science and Technology of Japan.

\newpage

\begin{figure}
\caption{%
(a) Specific heat, $C_p$, (b) magnetic susceptibility, $M / H$, 
and (c) electrical resistivity, $\rho$, for the Na$_{0.75}$CoO$_2$ sample 
with respect to temperature. 
The inset in Fig.~(c) shows the $\rho - T$ relation 
for a wider temperature range. 
}
\label{22K-anomalies}

\caption{%
Inverse susceptibility ($\chi^{-1}$) vs temperature relation 
at $H$ = 10 Oe for Na$_{0.75}$CoO$_2$. 
The solid line represents the best fit to the Curie-Weiss law 
above $T_m$ = 22 K. 
}
\label{CW-plot}

\caption{%
$M - H$ loops for the Na$_{0.75}$CoO$_2$ sample 
at 2 K, 10, 20, and 50 K. 
The inset shows a magnified $M - H$ loop at 2 K 
in the weak-field region. 
}
\label{M-H}

\caption{%
Spontaneous magnetization, $M_s$, of the Na$_{0.75}$CoO$_2$ sample 
as a function of temperature. 
The $M_s$ value was obtained at $H$ = 0 by linearly extrapolating 
the $M - H$ curve. 
}
\label{Ms-T}

\caption{%
Dependence of the degree of MR effect on temperature 
at a field of 7 T for the Na$_{0.75}$CoO$_2$ sample. 
The MR degree is defined as 
$\Delta \rho_H / \rho_0 \equiv (\rho_H - \rho_0) / \rho_0$ 
where $\rho_H$ is the electrical resistivity under an applied 
magnetic field ($H$) while $\rho_0$ is that at zero field. 
The inset shows the $(\Delta \rho_H / \rho_0) - H$ relation 
at various temperatures. 
Note that the $\Delta \rho_H / \rho_0$ values 
(below $T_m$) are all positive.
}
\label{MR}

\caption{%
Kohler's plot for the Na$_{0.75}$CoO$_2$ sample 
at various temperatures below $T_m$ = 22 K. 
Note that a temperature-independent relationship is established 
between $\Delta \rho_H / \rho_0$ and $(H / \rho_0)^2$, 
which indicates that the positive MR effect of Na$_{0.75}$CoO$_2$ 
is most likely originated from the conventional orbital motion of carriers. 
}
\label{Kohler}
\end{figure}


\begin{thebibliography}{00}
\bibitem{comment2}
It can be seen that a significant $H$-linear component exists 
in the $M-H$ plots even below $T_m$ (see Fig.~\ref{M-H}). 
This suggests that only a small fraction of cobalt spins contributes 
to the magnetic ordering, giving rise to the spontaneous magnetization. 
Here we subtracted the 2 K $H$-linear component from the magnetization 
data at any temperatures, and thus constructed Arrott plots again. 
This is because the Arrott plots should work only for the ferromagnetic 
portion. 
In the ``modified'' Arrott plots, a distinct change in the curvature of 
the $M^2$ vs $H/M$ curve was recognized about $5 - 10$ K. 
This is consistent with our model concerning 
the magnetic behavior of Na$_{0.75}$CoO$_2$.
\bibitem{comment1}
It was suggested that below $T_m$ the Fermi surface for the 
higher-mobility carriers suppresses the increase in resistivity 
caused by the pseudogap opening. 
That is the reason why a kink appears for Na$_{0.75}$CoO$_2$ 
instead of a hump as for classical SDW systems, i.e. Cr metal.
\end{thebibliography}
\end{document}